\documentclass[sigconf]{acmart}

\usepackage{hyperref}
\usepackage{balance}
\usepackage{graphicx}

\AtBeginDocument{%
  \providecommand\BibTeX{{%
    \normalfont B\kern-0.5em{\scshape i\kern-0.25em b}\kern-0.8em\TeX}}}
\copyrightyear{2020}
\acmYear{2020}
\setcopyright{acmlicensed}\acmConference[ICSEW'20]{IEEE/ACM 42nd International Conference on Software Engineering Workshops }{May 23--29, 2020}{Seoul, Republic of Korea}
\acmBooktitle{IEEE/ACM 42nd International Conference on Software Engineering Workshops (ICSEW'20), May 23--29, 2020, Seoul, Republic of Korea}
\acmPrice{15.00}
\acmDOI{10.1145/3387940.3392210}
\acmISBN{978-1-4503-7963-2/20/05}

\begin{document}

\title{Mutant Density: A Measure of Fault-Sensitive Complexity}

\author{Ali Parsai}
\email{ali.parsai@flandersmake.be}
\orcid{0000-0001-8525-8198}
\affiliation{
  \institution{Flanders Make}
  \streetaddress{Oude Diestersebaan 133}
  \city{Lommel}
  \country{Belgium}
}

\author{Serge Demeyer}
\email{serge.demeyer@uantwerpen.be}
\orcid{0000-0002-4463-2945}
\affiliation{
  \institution{University of Antwerp and Flanders Make}
  \streetaddress{Middelheimlaan 1}
  \city{Antwerp}
  \country{Belgium}
}

\begin{abstract}
 Software code complexity is a well-studied property to determine software component health. However, the existing code complexity metrics do not directly take into account the fault-proneness aspect of the code. We propose a metric called mutant density where we use mutation as a method to introduce artificial faults in code, and count the number of possible mutations per line. We show how this metric can be used to perform helpful analysis of real-life software projects.
\end{abstract}




\maketitle

\section{Introduction}

The software has grown from a niche afterthought to an all encompassing aspect of products in many industries. In addition, software products are becoming more and more complex everyday. Ensuring quality of the software is therefore becoming more important and more difficult at the same time. This makes the quality of the software an important aspect of its health~\cite{Hyrynsalmi2015}.

The complexity of the software is often tied to its quality. In particular, software complexity metrics are used as a predictor of its health and maintainability~\cite{Monteith2014,Kafura1987}. These metrics are used extensively in defect prediction literature as well (e.g. \cite{Meulen2007} and \cite{Gao2011}), and are generally known to be an indication of fault
-proneness of the code. Therefore they can be used as an indication for software developers to predict and avoid defects~\cite{Zhang2007}.

However, complexity metrics by themselves  are not tied to different defect types. While high complexity of a software component is a sign of potential for a fault, it does not clarify what parts of code are more likely to contain the fault. Therefore, a metric that  pinpoints statements more likely to be at risk of a fault is a useful tool for the developer in maintaining the software. Such a metric needs to take into account all possible ways a statement can become faulty according to a particular fault model.

Such a mechanism already exists in the field of mutation testing. Mutation testing is a process in which a fault is deliberately introduced in a  software component and then the tests are executed to see whether they are able to detect the fault. The faulty version of the software in this method is called a \textit{mutant}. The mutants are created based on fault models that aim to replicate common mistakes of the domain in which the test quality needs to be measured, hence the mutants are sensitive to the type of faults that are intended to be caught~\cite{Jia2011,Papadakis2018}. 

In this article, we leverage the mentioned property of mutants to create a new fault-sensitive metric for complexity. For this, we count the number of mutants that can be generated for each line of code. We call this metric \textit{mutant density}. We showcase two types of analysis by using mutant density. Through such analysis, developers can improve the quality of their software components, and consequently the health of their software.

This article is structured as follows:
In Section \ref{sec:background} we briefly describe the background information and related work in the context of this article.
In Section \ref{sec:md} we describe mutant density metric in detail.
In Section \ref{sec:analysis} we provide few ways this metric can be utilized.
Finally, in Section \ref{sec:conclusion} we conclude the article and present future research directions.

\section{Background}
\label{sec:background}
In this section, we briefly describe the background information and related work in the context of this article.

\subsection{Complexity Metrics}
Complexity metrics are used to quantify the complexity of a unit of software. McCabe cyclomatic complexity is a widely-used and yet controversial metric created by McCabe in 1976~\cite{McCabe1976}. This metric  measures the number of linearly independent paths through a unit of code. While its use has been advocated since 1980's in software maintenance~\cite{Kafura1987}, it is considered an inferior metric for this purpose in academic circles~\cite{Ebert2016}. Yet, it is still used extensively in academic case studies and in practice (e.g.~\cite{Antinyan2014,Nugroho2011}). 
It is known that developers tend to change their behavior in order to avoid hot spots indicated by such metrics~\cite{Staahl2019}.
As a result, code complexity is a useful metric in the context of the software health, and particularly software component health~\cite{Monteith2014} and defect prediction~\cite{Lajios2009}.

\subsection{Mutation Operators}
A \textit{mutation operator} is a pattern that changes part(s) of a software system in order to introduce a fault. The faulty version of the code produced in this manner is called a \textit{mutant}.
The traditional mutation operators  were first reported in King et al.~\cite{King1991} for FORTRAN77 programming language. These operators basic language elements such as arithmetic operators, conditional statements, etc.  In 1996, Offutt et al. show that a smaller set of mutation operators can produce a similarly capable test suite. This reduced set of operators  remains popular in most mutation testing tools to date.

A major branch of academic research in mutation operators is focused on inventing new mutation operators to target emerging fault patterns and new programming paradigms such as targeting certain security problems~\cite{Shahriar2008a,Zeng2009}, language specific mutation operators~\cite{Abraham2009,Bradbury2006a,Silva2012,Parsai2018,Parsai2019}, or object-oriented mutation operators~\cite{Ma2002,Chen2006}.

Artificial faults created by mutants are known to be a good replacement to real faults in software experiments~\cite{Just2014,Andrews2005}. The mutation score also correlates well with defect density~\cite{Tengeri2016a}.

\subsection{LittleDarwin}
LittleDarwin is a mutation testing tool designed for Java source code. LittleDarwin has been used in several studies, and is capable of performing mutation testing on complicated software systems~\cite{Parsai2016,Parsai2019}. 
LittleDarwin contains two distinct set of mutation operators: traditional and null-type. For more information about LittleDarwin and its structure please refer to Parsai et al.~\cite{Parsai2017}. LittleDarwin is open source software, and the latest version can be found at \url{https://littledarwin.parsai.net/}.

\section{Mutant Density}
\label{sec:md}
\begin{figure}
\centering
\includegraphics[width=\linewidth]{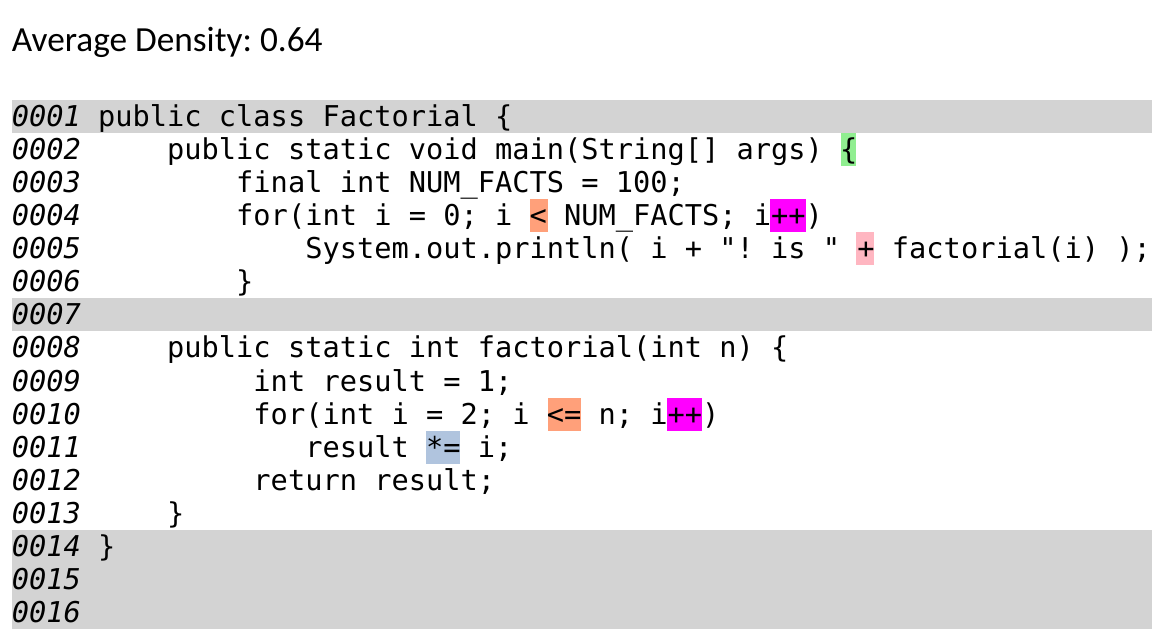}
\caption{An Example of the Usage of Mutant Density}
\label{fig:example}
\end{figure}

Mutant density is defined as the number of mutants that are generated for each line of code. Average mutant density for a source file is then defined as the sum of mutant density of each \textit{relevant} line divided by the number of lines of code in the file. For the calculation of this metric, we only consider non-blank lines of code within methods and constructors relevant.  Figure~\ref{fig:example} shows a mutation report where mutant density is calculated. In this figure, the colored highlights show mutants, the white background means the line is relevant, and gray background means non-relevant. For example, \texttt{for(int i = 0; i < NUM\_FACTS; i++)} has a mutant density of two, since \texttt{i < NUM\_FACTS} can be mutated to  \texttt{i >= NUM\_FACTS} and \texttt{i++} can be mutated to \texttt{i--}.

It is apparent that this metric is subject to change based on the fault model that the mutation engine uses. For example, using a different set of mutation operators, more mutants can be generated from \texttt{i < NUM\_FACTS}, perhaps changing the constant to zero, or changing the '<' operator to '<=' and '=='. This, however, is not a bug, but rather a feature of this metric: it allows the developer to fine-tune the fault model to measure what matters in the particular context of their project rather than rely on the generic fault models.

\begin{figure*}
\centering
\includegraphics[width=\linewidth]{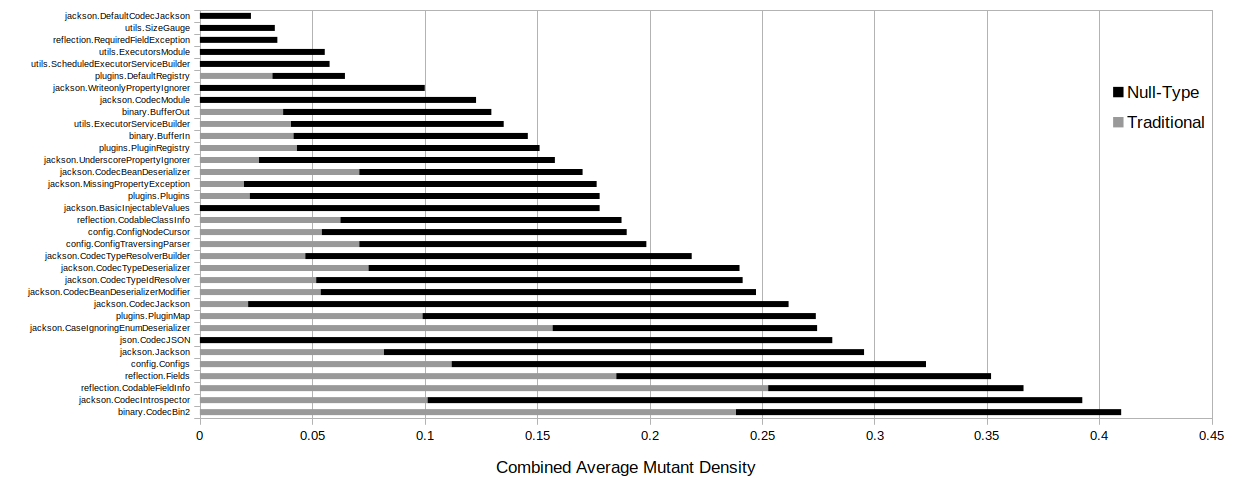}
\caption{Combined Average Mutant Density for Compilation Units in AddThis Codec}
\label{fig:combined-density}
\end{figure*}

\section{Analysis}
\label{sec:analysis}
For the purposes of this article, we bring examples of the use of mutant density metric from  real-life open source Java  project AddThis Codec\footnote{\url{https://github.com/addthis/codec}}. What follows is two sample analyses that we provide using this metric. We use LittleDarwin to calculate mutant density for traditional and null-type faults.

\subsection*{Finding Fault-Prone Components}

Using average mutant density metric at compilation unit level allows us to visualize the complexity of each unit with regards to the used fault model. Therefore it helps the developer in locating fault-prone components, and to improve the quality of the code.

Figure \ref{fig:combined-density} shows the combined average mutant density for compilation units in AddThis Codec. In this Figure, the gray bars show the value of average mutant density based on traditional mutation operators, and black bars  show the value of average mutant density based on null-type mutation operators for each compilation unit. 
Here, we can see that \texttt{binary.CodecBin2} and \texttt{jackson.CodecIntrospector} are the two most complex compilation units in this project. In addition,   \texttt{binary.CodecBin2} is more prone to traditional faults (such as mistakes in arithmetic and conditional operators) than null-type faults. An interesting observation is the fact that \texttt{json.Codec.JSON} is not really prone to traditional faults, since it lacks those mutable structures, however, it is highly susceptible to null-type faults.

\subsection*{Rewriting Complex Statements}

\begin{figure*}
\centering
\includegraphics[width=0.65\linewidth]{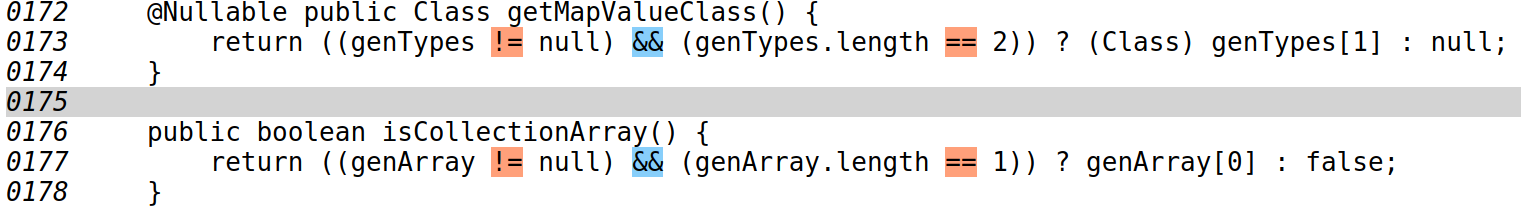}
\caption{An Example of Fault-Prone Statements in \texttt{reflection.CodableFieldInfo} to be Rewritten}
\label{fig:rewriting}
\end{figure*}

By visualizing mutant density of each line in a compilation unit, it becomes instantly apparent where the complexity lies. Using this information, a developer can quickly refactor and rewrite high-complexity statements and reduce the chance for future mistakes.
 
In Figure~\ref{fig:rewriting}, two methods from AddThis Codec located in compilation unit \texttt{reflection.CodableFieldInfo} are shown. Both these methods consist of a single return line where all the logic of the method is performed. Using mutant density one can instantly see the complexity of this statement is high. As a solution, these return statements can be separated into several lines of code, and by doing so, one can increase the readability and decrease fault-proneness of these statements.

\section{Conclusion and Future Work}
\label{sec:conclusion}
Code complexity is often used as a surrogate metric for fault-proneness of software components, however, the existing code complexity metrics do not directly aim at fault-proneness  aspect of the code. In this article, we proposed the use of a new metric called mutant density that uses a customizable fault model to calculate complexity of each line of code. We show how this metric can be used by developers to extract useful insight into the health of their software components.

The research into mutant density can be continued in many future directions. The relation between mutant density and defect density can show whether the previous assumptions about the quality of artificial faults holds true. The customization of the fault-model allows for several types of mutation operators to be tried out. The current set of mutation operators are optimized for increasing the quality of software tests. Therefore, it is possible that a different set of mutation operators is required to produce more accurate mutant density calculations. Finally, the effects of redundant and equivalent mutants on this metric are unknown at this time and demand further investigation.

\begin{acks}
This work is sponsored by:\\ (a) ITEA$^3$ \textsf{TESTOMAT Project} (number 16032), sponsored by VINNOVA -- Sweden's innovation agency;\\
(b) Flanders Make vzw, the strategic research centre for manufacturing industry.
\end{acks}

\bibliographystyle{ACM-Reference-Format}
\balance
\bibliography{ParsaiSOHEAL2020}

\end{document}